\documentclass[thmsa,11pt,a4paper,oneside,final,ukenglish]{article}
\usepackage{graphicx}

\begin{document}

\title{Scaling behavior in the heart rate variability characteristics with
age}
\author{Isabel M. Irurzun,$^{1)}$, Magdalena M. Defeo,$%
^{2)}$,L. Garavaglia$^{1)}$,J. Thomas Mailland$^{1)}$,
E. E. Mola$^{3)}$\\
\\
(1) CCT La Plata- CONICET. Instituto de Investigaciones Fisicoqu\'{i}micas
Te\'{o}ricas y Aplicadas \\
(INIFTA), Facultad de Ciencias Exactas, Universidad Nacional de La Plata. \\
La Plata, Rep\'{u}blica Argentina. \\
(2) Hospital Interzonal General de Agudos ''Prof. Dr. RodolfoRossi''\\
La Plata, Rep\'{u}blica Argentina. \\
(*) Corresponding author: i\_irurzun@hotmail.com.ar
(3) In memory}
\date{}
\maketitle
\begin{abstract}
In this work we study the characteristics of the heart rate variability
(HRV) as a function of age and gender. The analyzed data include previous
results reported in the literature. The data obtained in this work
expand the range of age studied until now revealing new behaviors
not reported before. We analyze some measurements in the time domain,
in the frequency domain and nonlinear measurements. We report scaling
behaviors and abrupt changes in some measurements. There is also a
progressive decrease in the dimensionality of the dynamic system governing
the HRV, with the increase in age that is interpreted in terms of
autonomic regulation of cardiac activity.
\end{abstract}

\section{Introduction}

Heart rate variability (HRV) is the physiological variation in the
duration of cardiac cycles {[}1,2{]}. With the development of electrocardiographic
devices the term was related to the variation in the duration of the
RR-intervals in an electrocardiographic record. The HRV is mainly
controlled by the autonomic nervous system (ANS) through the interplay
of sympathetic and parasympathetic neural activity mainly at the sinus
node {[}3,4{]}. In general, the HRV is influenced by many several
factors such as chemical, hormonal and neural modulations, circadian
changes, exercise, emotions, posture and preload. The adaptation of
the heart rate to changing factors is carried out by the activity
of different regulatory subsystems, i.e. activity of vasomotor and
respiratory centers, of baroreflex and chemoreflex closed loop regulation,
of cardiovascular reflexes mediated by vagal and sympathetic afferences,
and of vascular and thermoregulation. The variety of regulatory
subsystems results in a complex linear and nonlinear temporal behavior,
which changes with age and pathologic conditions. Several studies demonstrated
age-related and gender-related variation in long-term HRV characteristics.
It was reported that autonomic activities diminish with age in both
genders and that gender-related variation in parasympathetic regulation
decreases after the age of 50 years {[}5-11{]}.

HRV characteristics were proposed as predictors of the risk of premature
mortality after myocardial infarction or development of congestive
heart failure, diagnosis of autonomic dysfunction in diabetes, non-invasive
estimation of the autonomic modulation of the cardiovascular system
during stress, relaxation or the assessment of the effects of physical
training on fitness level. All of these are the reasons why the interest
in HRV is growing both in clinical and physiological studies {[}12-19{]}. 

Many mathematical methods to compute the HRV characteristics have
been developed\textemdash they may be grouped into statistical, spectral,
graphical, nonlinear, complexity, or information based{[}20-23{]}. 

In summary, there has been a huge effort from the world scientific and medical
community to have reliable measurements of the HRV characteristics
in normal and pathological conditions. Concerning the relationship
of HRV with gender and age, very extensive and complete studies can
be found in {[}2, 24{]} and references therein, which both together constitute
the broadest study of the HRV relationship with age we know.

In this work \textit{we expand the range of age studied in the literature
and reveal new behaviors that had not been detected until now}. We
include data from previous studies and show that our data are consistent
with them. We also analyze differences by gender. Our study is limited
to the study of some of the existing measurements, but the results show
the need to reanalyze all the others, and a further analysis (with
additional insights into the treatment of data) will be presented later.

This work is organized as follows: in the next section methodological
details are explained. They are equal to those also used in {[}21,22{]}
and comparable to the methodology used in {[}2, 24{]}.

Section 3 shows the results first as a function of age and then
distinguishing among genders. Dependences are rationalized adjusting
power law behaviors.

Finally we summarize our conclusions in Section 4.

\section{Procedure}

Holter recordings from healthy subjects were collected from volunteers
after an exhaustive interview and clinical examination. Those individuals without clinical symptoms of disease, without medication
and with electrocardiograms (ECG) within normal parameters according
to the criteria summarized in Table 1 were included{[}25,26{]}.

\begin{table}
\begin{tabular}{|c|c|}
\hline 
I & Minimal nighttime frequency>60/min\tabularnewline
\hline 
\hline 
II & Nighttime pauses<3 seg.\tabularnewline
\hline 
III & Ventricular extrasystoles < 100/24h, without couplets, bursts or polymorphism.\tabularnewline
\hline 
IV & Supraventricular extrasystoles < 100/24h, without bursts.\tabularnewline
\hline 
V & Absence of blocks or conduction disturbances.\tabularnewline
\hline 
\end{tabular}

\caption{\label{tab:normallity}Normality criteria for all Holters recorded
in the present work.}
\end{table}

Holters were recorded for 24 h with digital three-channel DMS300
7 and DMS300 3A recorders, and Galix recorders, using 3M electrodes{[}27{]}.
The automatically detected and classified electrocardiographic recording
events were examined and corrected by two cardiologists, and the artifacts
were removed as aforementioned. 

We applied quality criteria established in {[}25,26{]} to the all
time series used in the present work. Also stationarity was evaluated
and surrogate analysis was performed as in {[}28-31{]}.

Time series of a total of 195 healthy individuals were finally analyzed
(13 time series were taken from {[}32{]}, and 28 from {[}33{]}). They
are from 0 to 74 years old and 50\% of them are females. 

In Results we also introduced data from {[}2,24{]} for comparison purposes.
In total, data of about 500 healthy subject aged between 1 month and
99 years were evaluated.

\section{Results}

\subsection{Linear analysis}

The following linear indexes in the time domain were calculated: the
RR interval mean value, $<RR>$, the standard deviation, $SD_{RR}$,
the square root of the mean of the sum of the squares of differences
between consecutive RR intervals , $rMSSD_{RR}$, and the percentage
of the intervals that vary more than 50 ms from the previous interval,
$pNN50$. 

Figure 1 shows their dependence on age, including data from {[}2{]}
and {[}24{]}. Data from different sources show a good agreement among
them validating the general treatment of the measurements. Our data
expand the experimental range of age revealing unknown tendencies.
Indeed while $<RR>$ exhibits a monotonic behavior, $SD_{RR}$, $rMSSD_{RR}$
and $pNN50$ show an abrupt change at the age of 12 years not detected
so far. We rationalized Figure 1 results through scaling laws as
follows:

\begin{equation}
<RR>=(515\pm2)x^{(0.117\pm0.003)}
\end{equation}

\begin{equation}
SD_{RR}=\left\{ \begin{array}{c}
(79.4\pm0.8)x^{(0.25\pm0.01)}\begin{array}{ccccc}
 &  &  &  & x<12\end{array}\\
(295\pm12)x^{(-0.22\pm0.03)}\begin{array}{ccccc}
 &  &  &  & x>12\end{array}
\end{array}\right\} 
\end{equation}

\begin{equation}
rMSSD_{RR}=\left\{ \begin{array}{c}
(18,6\pm0.2)x^{(0.34\pm0.02)}\begin{array}{ccccc}
 &  &  &  & x<12\end{array}\\
(166\pm11)x^{(-0.46\pm0.04)}\begin{array}{ccccc}
 &  &  &  & x>12\end{array}
\end{array}\right\} 
\end{equation}

\begin{equation}
pNN50=\left\{ \begin{array}{c}
(0.037\pm0.001)x^{(0.78\pm0.07)}\begin{array}{ccccc}
 &  &  &  & x<12\end{array}\\
(5\pm1)x^{(-1.1\pm0.1)}\begin{array}{ccccc}
 &  &  &  & x>12\end{array}
\end{array}\right\} 
\end{equation}

where $x$ is the age in years.

Though the power law adjustments for ages above 12 years are statistically
worse than those for ages below 12 years, they are still better than or
equal to other linear or quadratic adjustments performed on the
same sets. 

\begin{figure}
\begin{centering}
\includegraphics{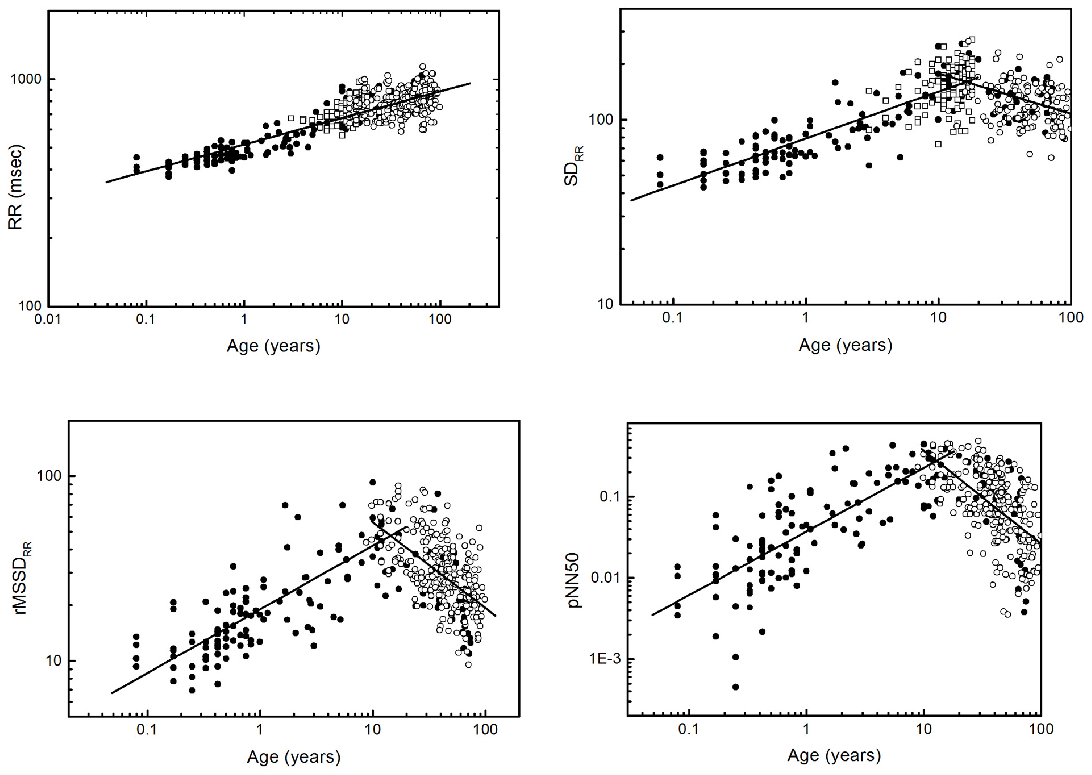}
\par\end{centering}
\caption{\label{fig:1-aging effect}Age dependence of some statistical indexes.
Filled circles are the data of this work, open squares are data from
{[}2{]}, and open circles are data from {[}24{]}.}
\end{figure}

Gender differences are shown in Figure 2, and Table 2 summarizes the
power law parameters in each case. For ages below 12 years there are
no significant differences with gender, while for ages above 12 years,
a slight but significant difference appears in $SD_{RR}$ and $pNN50$
(see Table 2)

\begin{figure}
\begin{centering}
\includegraphics{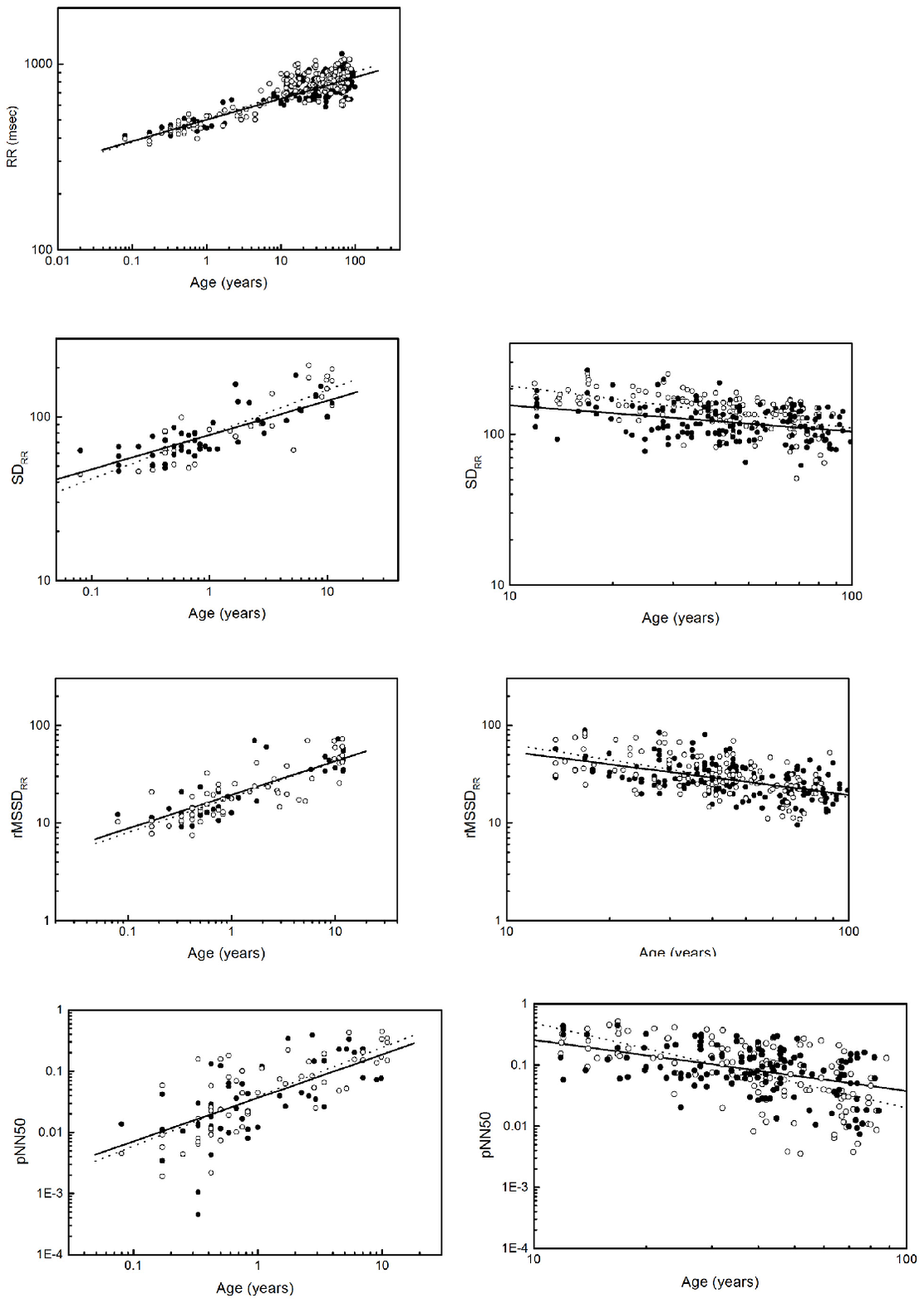}
\par\end{centering}
\caption{\label{fig:1-gender effect-1}Gender effect on different statistical
indexes. Filled circles are female subjects and open circles are male
subjects. Panels on the left show the adjustments for ages below 12, while
panels on the right show the adjustments for ages above 12. The adjustment
parameters are shown in Table 2. Solid lines correspond to female
subjects and broken lines correspond to male subjects.}
\end{figure}

\begin{table}
\begin{tabular}{|c|c|c|}
\hline 
Index & Male & Female\tabularnewline
\hline 
\hline 
$<RR>$ & %
\begin{tabular}{|c|c|}
\hline 
Eq. & $506(4)x^{0.125(5)}$\tabularnewline
\hline 
\hline 
N & $182$\tabularnewline
\hline 
R & $0.94$\tabularnewline
\hline 
p< & $10^{-4}$\tabularnewline
\hline 
\end{tabular} & %
\begin{tabular}{|c|c|}
\hline 
Eq. & $500(3)x^{0.115(4)}$\tabularnewline
\hline 
\hline 
N & $213$\tabularnewline
\hline 
R & $0.95$\tabularnewline
\hline 
p< & $10^{-4}$\tabularnewline
\hline 
\end{tabular}\tabularnewline
\hline 
$SD_{RR}$ & %
\begin{tabular}{|c|c|c|}
\hline 
 & below 12 & above 12\tabularnewline
\hline 
\hline 
Eq. & $77.6(8)x^{0.27(3)}$ & $398(4)x^{-0.28(4)}$\tabularnewline
\hline 
N & $57$ & $123$\tabularnewline
\hline 
R & $0.86$ & $-0.55$\tabularnewline
\hline 
p< & $10^{-4}$ & $10^{-4}$\tabularnewline
\hline 
\end{tabular} & %
\begin{tabular}{|c|c|c|}
\hline 
 & below 12 & above 12\tabularnewline
\hline 
\hline 
Eq. & $77.6(8)x^{0.21(3)}$ & $229(2)x^{-0.17(4)}$\tabularnewline
\hline 
N & $50$ & $148$\tabularnewline
\hline 
R & $0.80$ & $-0.36$\tabularnewline
\hline 
p< & $10^{-4}$ & $10^{-4}$\tabularnewline
\hline 
\end{tabular}\tabularnewline
\hline 
$rMSSD_{RR}$ & %
\begin{tabular}{|c|c|c|}
\hline 
 & below 12 & above 12\tabularnewline
\hline 
\hline 
Eq. & $19.5(8)x^{0.36(3)}$ & $200(20)x^{-0.54(7)}$\tabularnewline
\hline 
N & $58$ & $125$\tabularnewline
\hline 
R & $0.83$ & $-0.57$\tabularnewline
\hline 
p< & $10^{-4}$ & $10^{-4}$\tabularnewline
\hline 
\end{tabular} & %
\begin{tabular}{|c|c|c|}
\hline 
 & below 12 & above 12\tabularnewline
\hline 
\hline 
Eq. & $18.2(8)x^{0.34(4)}$ & $160(20)x^{-0.45(6)}$\tabularnewline
\hline 
N & $55$ & $143$\tabularnewline
\hline 
R & $0.79$ & $-0.51$\tabularnewline
\hline 
p< & $10^{-4}$ & $10^{-4}$\tabularnewline
\hline 
\end{tabular}\tabularnewline
\hline 
$pNN50$ & %
\begin{tabular}{|c|c|c|}
\hline 
 & below 12 & above 12\tabularnewline
\hline 
\hline 
Eq. & $0.038(2)x^{0.8(1)}$ & $12(2)x^{-1.4(2)}$\tabularnewline
\hline 
N & $58$ & $124$\tabularnewline
\hline 
R & $0.76$ & $-0.63$\tabularnewline
\hline 
p< & $10^{-4}$ & $10^{-4}$\tabularnewline
\hline 
\end{tabular} & %
\begin{tabular}{|c|c|c|}
\hline 
 & below 12 & above 12\tabularnewline
\hline 
\hline 
Eq. & $0.037(2)x^{0.7(1)}$ & $1.7(1)x^{-0.8(1)}$\tabularnewline
\hline 
N & $53$ & $133$\tabularnewline
\hline 
R & $0.67$ & $-0.43$\tabularnewline
\hline 
p< & $10^{-4}$ & $10^{-4}$\tabularnewline
\hline 
\end{tabular}\tabularnewline
\hline 
\end{tabular}

\caption{\label{tab:statistics}Power law adjustments by gender and age range
(where $x$ is the age). The numbers parentheses indicate
the error in the parameters, N is the number of data, R is the correlation
coefficient and p is the t-Student parameter. }
\end{table}

\subsection{Frequency domain measurements}

Heart rate variability time series exhibit power law behavior in the
frequency (1/beat) domain, which is manifested in the power spectrum
behavior and expressed as

\begin{equation}
S(f)\alpha f^{\beta}
\end{equation}

$\beta$ values were determined in this work by averaging the power
spectra of successive time series segments of 4096 beats. An example
is shown in Fig. 3. The procedure allows the elimination of high frequency
fluctuations and is detailed in {[}26{]}. Other frequency domain measurements
were defined such as the low frequency (LH) and high frequency (HF)
indexes and their ratio (LF/HF), etc. They will be discussed in a
further work in comparison with other short term measurements.

Figure 4 shows the dependence of $\beta$ on age. A nonmonotonic
behavior is observed which was no reported so far: an increase of
$\beta$ values appears at the interval extremes, where the action
of each one of the subsystems of the ANS dominates (either sympathetic
or parasympathetic tones). Also a minimum at the age of 1 year, is
revealed which was not reported so far and deserves to be further explored.

\begin{figure}
\begin{centering}
\includegraphics{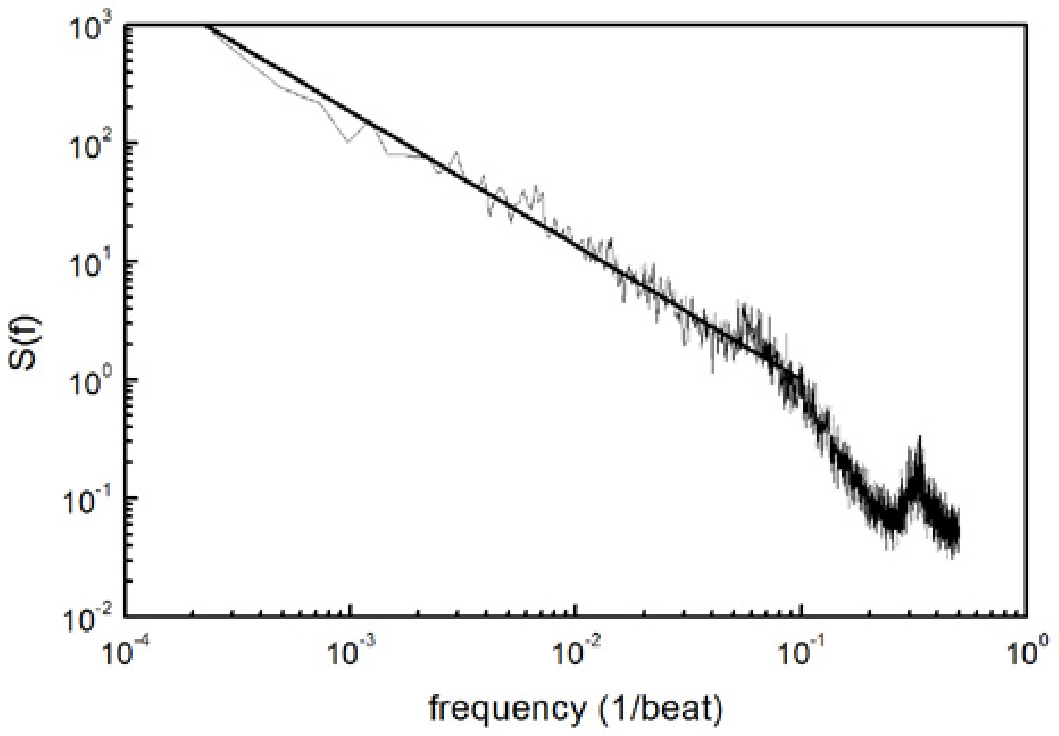}
\par\end{centering}
\caption{Averaged power spectrum of a healthy adult subject and linear adjustment
at low frequencies to determine $\beta$. For details see {[}26{]}}
\end{figure}

\begin{figure}
\begin{centering}
\includegraphics{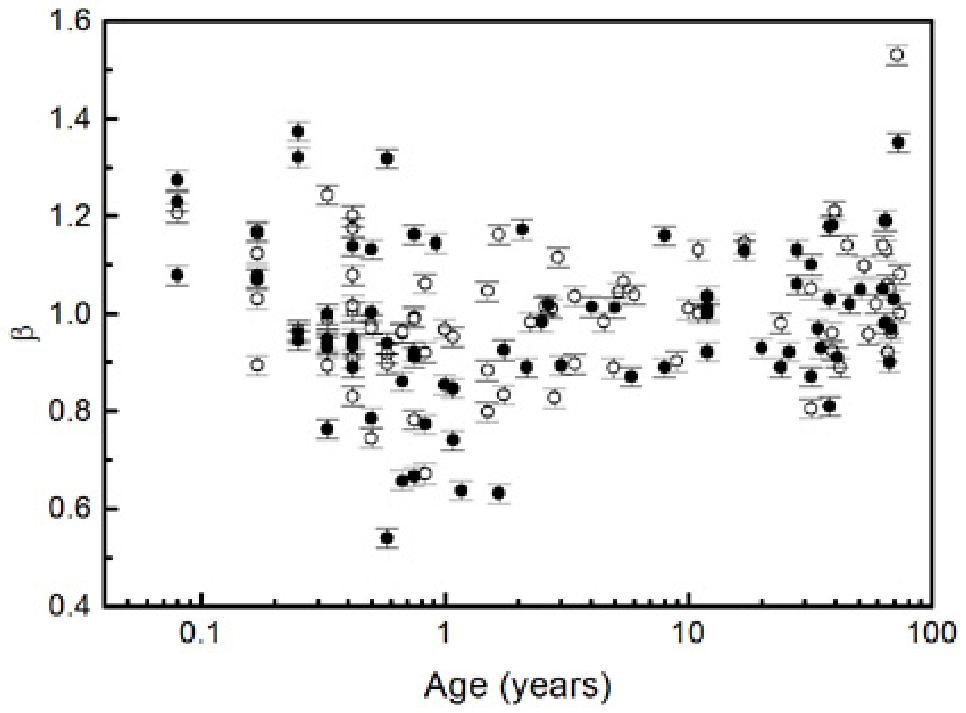}
\par\end{centering}
\caption{$\beta$ dependence on age and gender. Filled circles are female subjects
and open circles are male subjects}
\end{figure}

\subsection{Nonlinear analysis}

Heart rate variability time series can be thought as a sequence of
observations ${s_{n}}$ performed on a multidimensional dynamic system. 

To unfold the multidimensional structure of the system by using a
scalar sequence ${s_{n}}$ of data, the method of delays is employed
in nonlinear sciences. 

Vectors in an embedding space are formed from time-delayed values
of the scalar measurements $s_{n}=(s_{n-(m_{0}-1)\tau},s_{n-(m_{0}-2)\tau},\ldots,s_{n})$
, where $m_{0}$ is the minimal embedding dimension and $\tau$ is
the delay time. Both $m_{0}$ and $\tau$ provide fundamental information
on the dynamic system;$m_{0}$ gives the dimension to completely unfold
the trajectory of the system in the phase space. 

The false nearest neighbor (FNN) method was proposed by Kennel et
al. to determine $m_{0}$ {[}34{]}. The idea is very intuitive and
is based on the fact that in the embedding dimension ($m_{0}$), the
trajectories or the attractor reconstructed from a physical observable
by the delayed coordinate method are a biunique image of the attractor
in the original phase space. In particular, the topological characteristics
of the attractor are preserved, despite the changes in the radii of
curvature, the trajectories and the radius of the neighborhood of
a point (according to Lyapunov exponents). Hence, the false nearest
neighbor method consists of reconstructing the attractor in progressively
greater$m<m_{0}$ dimensions and determining the number of average
neighbor points (within a neighborhood of radius $\varepsilon$ )
of each of the attractor points. As the topologies of the original
attractor projections are not necessarily preserved in the reconstructed
attractor, a point belonging to a neighborhood for a given m value
may belong to another neighborhood for a greater $m$ value. We will
then say that in our neighborhood that point was a false nearest neighbor.

The method consists of calculating the false nearest neighbor fraction
for progressively larger values of $m$ . When $m=m_{0}$ , the false
nearest neighbor fraction should stabilize and ideally take a zero
value. In practice, we determined $m_{0}$ as the value of $m$ where
the false nearest neighbor fraction curve stabilizes, i.e, that the
absolute difference between two successive values is less than$0.0005$.

Eventually, the result also depends on both the length and the delay
time $\tau$ of the time series, and an adequate comparison of the
results will require a careful evaluation of the algorithm and the
standardization of the procedures used. In this study we used the
nearest neighbor algorithm provided by the TISEAN software package
{[}35,36{]}. We chose a value of $\tau=1$ for all the time series.
This value, as well the algorithm
performance as a function of the length of the time series, has been tested previously in our work.

The $FNNF10$ index is just the nearest neighbor fraction for $m=10$
(regardless of the $m_{0}$ value). This magnitude may be taken as
a measure of the error in the reconstruction of the attractor for
that value of $m$. 

Figure 5 shows the dependence of $m_{0}$ on age and gender.

\begin{figure}
\begin{centering}
\includegraphics{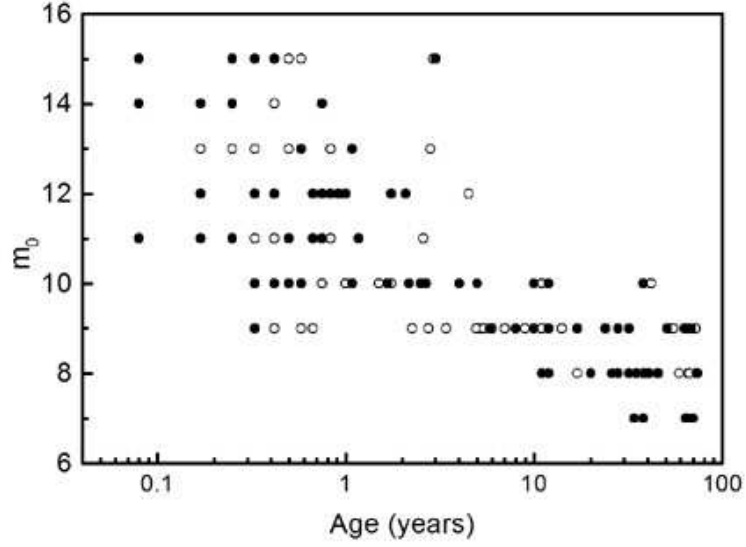}
\par\end{centering}
\caption{$m_{0}$ dependence on age and gender. Filled circles are female subjects
and open circles are male subjects}
\end{figure}

Figure 6 shows the dependence of FNNF10 on age for different genders.
The results were rationalized adjusting scaling equations as follows:

\begin{equation}
FNNF10=0.065(3)x^{-0,75(5)}
\end{equation}

\begin{equation}
FNNF10=0.072(4)x^{-0,79(5)}
\end{equation}

where x is the age and Eq (6) is valid for males while Eq (7) is valid
for females.

\begin{figure}
\begin{centering}
\includegraphics{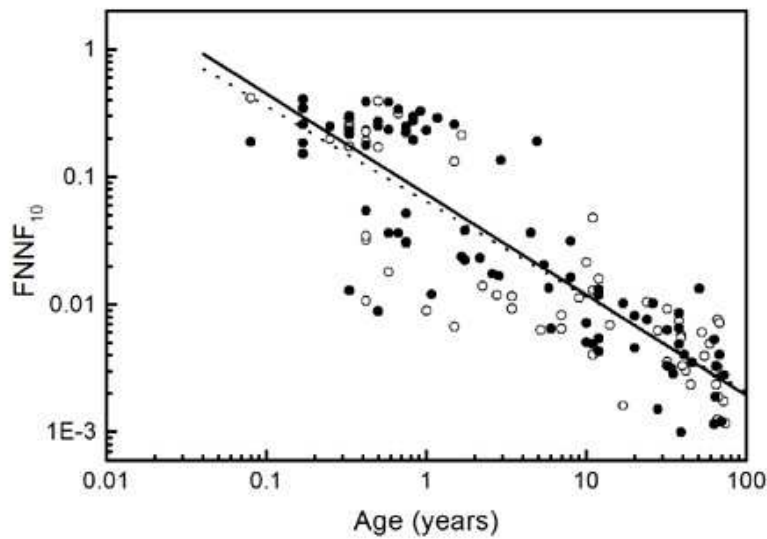}
\par\end{centering}
\caption{$FNNF_{10}$ dependence on age and gender. Filled circles are female
subjects and open circles are male subjects. The broken line corresponds
to Eq. (6) and the solid line to Eq. (7).}
\end{figure}

\section{Discussion and Conclusions }

Data analyzed in the present work are of different sources and correspond
to about 500 healthy subjects covering the most large range of ages
analyzed so far. In the present work 195 records were acquired. There
are many other studies based on age and gender in the literature.
Although the results reported by them are consistent with ours, a
direct comparison was not possible because of the way in which the
data were reported. The age of the individuals should be considered
as a continuous variable to detect the functional dependences as
reported in this work.

The main conclusions are:

$<RR>$follow a scaling relationship with age that is independent
of gender.

Statistical measures such as $SD_{RR}$, $r-MRSSD_{RR}$ and $pNN50$
show an abrupt change at the age of 12 years. We assume the same cutoff
(independent of gender) in all cases for simplicity, but this ansatz
should be further studied. Below 12 years, the results are independent of
gender, while above 12 years there seems to be a slight dependence on gender.

Other statistical measures also deserve to be explored.

Two previously developed non linear measurements, $m_{0}$ and $FNNF10$,
were also studied as a function of age and gender. $m_{0}$ is the
minimum number of topological dimensions to unfold the dynamical system
governing the HRV. It is highly variable among individuals but it
is always higher than 9 for ages below 10 years and always lower than
10 for ages up to 10. $FNNF10$ is the nearest neighbor fraction for
$m=10$ (regardless of the $m_{0}$ value). This magnitude diminishes
with age also following a scaling behavior that is independent of
gender. The decrease of $FNNF10$ is consistent with the fact that $m_{0}$
takes values that for children are higher than those for adults.
One could relate the behavior of the dynamic system to the changes
in the autonomic modulation of HRV. 

The autonomic activity diminishes with age in both genders, and the dynamic system evolves in a topological space of the decreasing
dimension, i.e, with a progressively lower number of dynamic variables
influencing the HRV.

These changes would also be reflected in the dependence of $\beta$
on age, but further studies are necessary to reveal them and they will
be presented later.

\section{Acknowledgments }

This research project was financially supported by the National Research
Council of Argentina (CONICET), the National University of La Plata;
and the ANPCyT.

\end{document}